\documentclass[final,3p,11pt,twocolumn]{elsarticle}

\usepackage{epsfig}
\usepackage{amssymb}
\usepackage{amsmath}
\usepackage{relsize}
\usepackage[T1]{fontenc}

\expandafter\let\csname equation*\endcsname\relax
\expandafter\let\csname endequation*\endcsname\relax 
\usepackage{amsmath}
\usepackage{amssymb}

\biboptions{compress}
\journal{Physics Letters A}

\def\vecb#1{\boldsymbol{#1}}

\def\ave#1{\langle#1\rangle}
\def\max#1{\{#1\}}
\def\dis#1{\langle\Delta^2#1\rangle}
\def\ske#1{\langle\Delta^3#1\rangle}
\def\={\!=\!}
\def\>{\!>\!}
\def\<{\!<\!}
\def\-{\!-\!}
\def\+{\!+\!}
\def\uvo#1{\lq\lq #1\rq\rq}
\def\ESQPT{\textsc{esqpt}}
\def\ESQPTs{\textsc{esqpt}s}

\def\QPTs{\textsc{qpt}s}
\def\DoF{\textsc{d}o\textsc{f}}
\def\DoFs{\textsc{d}o\textsc{f}s}
\def\NDSP{\textsc{ndsp}}

\begin{document}

%%%%%%%%%%%%%%%%%%%%%%%%%%%%%%%%%%%%%%%%%%%%%%%%%%%%%%%%%%%%%%%%%%%%%%%%%%%%%%%%%%%%%%%%%%%%%%%%%%%%%%%%%%%%%%%%%%%%%%%%%%%%%%%%
\begin{frontmatter}

\title{Heat capacity for systems with excited-state quantum phase transitions}

\author{Pavel Cejnar}
\author{Pavel Str{\'a}nsk{\'y}\corref{cor}} 
\address{Institute of Particle and Nuclear Physics, Faculty of Mathematics and Physics, Charles University, 
  V~Hole{\v s}ovi{\v c}k{\' a}ch 2, 180\,00 Prague, Czech Republic}
\cortext[cor]{The corresponding author; email address:\\ stransky@ipnp.troja.mff.cuni.cz}

\begin{abstract}
Heat capacities of model systems with finite numbers of effective degrees of freedom are evaluated  
using canonical and microcanonical thermodynamics. Discrepancies between both approaches, which are observed even in the infinite-size limit, are particularly large in systems that exhibit an excited-state quantum phase transition. The corresponding irregularity of the spectrum generates a singularity in the microcanonical heat capacity and affects smoothly the canonical heat capacity.
\end{abstract}

%\noindent{\it PACS numbers}: 05.30.Rt, 45.20.Jj, 64.60.Bd


\end{frontmatter}

\section{Introduction}
\label{sec:Int}
 
The heat capacity has served as a fundamental probe to the structure of matter already since the early days of statistical mechanics.
We may remind its role in the formation of quantum physics.
Being sensitive to the number and character of the microscopic degrees of freedom activated in the system at a given temperature, the heat capacity represents an important tool of theoretical, experimental and also applied research up to the present days \cite{Rei16}.

Thermal properties of condensed matter systems are mostly studied in the thermodynamic limit, in which the number of particles as well as the number of degrees of freedom (\DoFs) grow to infinity.
However, with the advent of experiments based on controlled quantum devices (like cold atom or quantum electrodynamical setups \cite{Gar16}), the thermodynamics of systems with limited numbers of effective \DoFs\ becomes relevant.
In such systems, the measure of size (e.g., number of particles) is not directly related to the number of relevant \DoFs, the latter including only a limited set of collective coordinates which play an active role in the given arrangement.
The infinite-size limit of such \uvo{finite} systems can have rather unusual thermodynamic properties.

An interesting phenomenon that occurs in the infinite-size limit of finite-\DoF\ systems is the excited-state quantum phase transition (\ESQPT).
It is a singularity of the density and flow of the quantum spectrum as a function of excitation energy and suitable control parameters \cite{Cej06,Cap08,Cej08,Rel08,Fer11,Bra13,Lar13,Die13,Bas14,Str14,Str15,Cej15,Kop15,Pue15,San15,San16,Str16}.
The \ESQPTs\ extend the ground-state quantum phase transitions at zero temperature (\QPTs) \cite{Sac99} and their possible links to thermal phase transitions were naturally considered.
Recent analyses \cite{Bas16,Rel16} performed within the Dicke model of single-mode superradiance showed that the thermal phase transition and \ESQPT\ spectral singularities present in that model are not directly connected (though interesting links exist).
It is however unclear to what extent this conclusion relies on the specific model employed, so the general problem of thermodynamic signatures of \ESQPTs\ persists.

In this work, we proceed in the study of this problem using the canonical and microcanonical heat capacity.
This quantity is a natural choice because of its ability to indicate thermal phase transitions.
We show that the presence of an \ESQPT\ in the spectrum does not generate any singularity of the canonical heat capacity, which basically confirms results of the previous analyses.
At the same time, we observe \ESQPT-related singularities in the microcanonical heat capacity and demonstrate dramatic discrepancies between canonical and microcanonical descriptions in the systems with \ESQPTs.

Here is a plan of the paper:
The formulas for canonical and microcanonical heat capacity are outlined in Sec.\,\ref{sec:CamiC}.
Heat capacities in presence of \ESQPTs\ for various \DoF\ numbers are analyzed in Sec.\,\ref{sec:EsqptCgen}. 
Numerical examples are discussed in Sec.\,\ref{sec:EsqptCnum}.
A summary is given in Sec.\,\ref{sec:Conc}.

\section{Canonical vs. microcanonical heat capacity}
\label{sec:CamiC}

From elementary statistical mechanics we know that descriptions based on the microcanonical, canonical and grandcanonical ensembles become equivalent in the thermodynamic limit, when the system becomes infinitely large, but away from that limit they provide rather different pictures \cite{Rei16}.
Microcanonical thermodynamics, which is based on the original Boltzmann definition of entropy, represents an adequate approach to the equilibrium state of virtually isolated systems with a finite (moderate) number of elementary constituents.
It is therefore considered as the most natural background for the theory of phase transitions in finite systems  \cite{Gro01,Dun06,Bro07,Kas08}.

Our present motivation for invoking canonical and microcanonical approaches is nevertheless different.
We show that concerning the validity of these approaches, the key question is not only whether the system contains an infinite number of particles, but also whether all these particles retain their independent degrees of freedom.
We focus on systems that can be called infinite and finite at the same time: {\em infinite\/} in the sense of an asymptotically large size parameter $N$, and {\em finite\/} for the overall number $f$ of effective \DoFs\ remains constant (preferably small) regardless of $N$.
It turns out that the canonical and microcanonical descriptions of such systems may give rather different predictions even in the (incomplete) \uvo{thermodynamic} limit $N\to\infty$.

Consider for example an $N_{\rm s}$-site array of trapped 2-level atoms (or spin-1/2 particles), each site being filled with $N_{\rm a}$ atoms. 
Assume that the initial state, intra- and inter-site interactions, and a thermalization mechanism (e.g. via an external field) are such that the square of the total quasispin of each site is conserved in its maximal value $N_{\rm a}/2$ (a fully symmetric state of all atoms involved).
This ensures a strongly collective behavior of the system even in presence of thermal fluctuations.
The dynamics of each site is described by a single Bloch sphere with radius $N_{\rm a}/2$ carrying a single collective \DoF.
So while the system's size is measured by $N=N_{\rm a}N_{\rm s}$, its overall \DoF\ number is $f\=N_{\rm s}$.
Since at temperature $T$ each site receives an overall thermal energy $T$ rather than $N_{\rm a}T$ (we set Boltzmann constant to unity), the thermal energy is expected to be extensive in $f$ rather than $N$.
The $N\to\infty$ condition is therefore not sufficient for the true thermodynamic limit, which is only achieved if both $N,f\to\infty$.
On the other hand, the $N_{\rm a}\to\infty$ limit enforces the classical behavior of the system since quantum fluctuations of the size $\hbar$ can be neglected on the Bloch sphere with an asymptotically increasing radius.  

These expectations can be verified by the behavior of heat capacity.
The canonical heat capacity is given by the familiar formula 
\begin{equation}
C^{\rm can}(\beta)=\frac{\partial\ave{E}_{\beta}}{\partial T}=\beta^2\frac{\partial^2\ln Z}{\partial\beta^2} 
\label{CanonC}
\,,
\end{equation}
where $\ave{E}_\beta$ is a thermal average of energy at an inverse temperature $\beta\=1/T$ and $Z(\beta)=\sum_ke^{-\beta E_k}=\int\rho(E)e^{-\beta E}dE$ stands for the canonical partition function.
Note that we assume a system with a discrete energy spectrum $E_k$ described by the level density $\rho(E)=\sum_k\delta(E\-E_k)$, where $\delta$ is the Dirac function.

Since, as indicated above, the $N\to\infty$ limit of finite-$f$ systems coincides with the semiclassical limit (for a more detailed discussion see Refs.\,\cite{Str14,Cej15}), the level density of an infinite system can be represented by the semiclassical expression
\begin{equation}
\rho(E)=(2\pi\hbar)^{-f}\int\delta\left(E\-H(\vecb{x})\right)\,d\vecb{x}
\label{leden}
\,,
\end{equation}
where $H(\vecb{x})$ with $\vecb{x}\equiv(\vecb{q},\vecb{p})$ stands for the classical Hamiltonian of the system as a function of coordinates $\vecb{q}\equiv(q_1,...,q_f)$ and momenta $\vecb{p}\equiv(p_1,...,p_f)$.
Also the semiclassical partition function is given by a phase--space integral
\begin{equation}
Z(\beta)=\left(2\pi\hbar\right)^{-f}\int e^{-\beta H(\vecb{x})}d\vecb{x}
\label{Parti}
\,,
\end{equation}
and the heat capacity reads as
\begin{equation}
C^{\rm can}(\beta)=\beta^2\dis{E}_\beta
\label{CanonC1}
\,,
\end{equation}
where $\dis{E}_\beta$ denotes dispersion of energy in the classical thermal ensemble at inverse temperature $\beta$. 
For a general quantity $W$, the thermal dispersion
\begin{equation}
\dis{W}_\beta\equiv\left\langle(W\!\-\ave{W}_\beta)^2\right\rangle_\beta\=\ave{W^2}_\beta\-\ave{W}_\beta^2
\quad
\label{disper}
\end{equation}
is determined by means of a $2f$-dimensional integration over the whole phase space:
\begin{equation}
\ave{W^n}_\beta=\frac{\int W(\vecb{x})^n e^{-\beta H(\vecb{x})}d\vecb{x}}{\int e^{-\beta H(\vecb{x})}d\vecb{x}}
\label{avepq}
\,.
\end{equation}

For a Hamiltonian of the familiar form
\begin{equation}
H=\frac{\vecb{|p|}^{2}}{2}+V(\vecb{q})
\label{Hstand}
\,,
\end{equation}
where $K\propto\vecb{|p|}^{2}$ is the quadratic kinetic energy and $V$ a potential energy, the canonical heat capacity reduces to
\begin{equation}
C^{\rm can}(\beta)=\frac{f}{2}+\beta^2\dis{V}_\beta
\label{CanonC2}
\,.
\end{equation}
Here the first term results from the evaluation of the momentum--space integrals involved in $\dis{K}_\beta$, while the second term is proportional to a thermal dispersion of the potential energy $\dis{V}_\beta$, which is obtained solely by integration over the configuration space using
%, via the following expression ($n\=1,2$):
\begin{equation}
\ave{V^n}_\beta=\frac{\int V(\vecb{q})^n e^{-\beta V(\vecb{q})}d\vecb{q}}{\int e^{-\beta V(\vecb{q})}d\vecb{q}}
\label{aveq}
\,.
\end{equation}

On the other hand, the microcanonical determination of the heat capacity makes use of the microcanonical inverse temperature
\begin{equation}
\beta^{\rm mic}(E)
%\equiv\frac{1}{T^{\rm mic}(E)}
=\frac{\partial\ln\rho}{\partial E}
\label{Microb}
\,.
\end{equation}
For a fixed value $\beta^{\rm mic}\=\beta$, this equation represents a condition for a vanishing derivative of the canonical thermal energy distribution 
\begin{equation}
w_{\beta}(E)=\frac{\rho(E)e^{-\beta E}}{Z(\beta)}
\label{Ether}
\end{equation}
at the given canonical inverse temperature $\beta$.
If this distribution has a single maximum, as expected in a typical situation, Eq.\,\eqref{Microb} determines its energy, which we denote $\max{E}_\beta$.
The microcanonical heat capacity is then defined as
\begin{equation}
C^{\rm mic}(\beta)\=\frac{\partial\max{E}_\beta}{\partial T}\=-\beta^2\left(\frac{\partial^2\ln\rho}{\partial E^2}\right)_{E=\max{E}_\beta}^{-1}
\label{MicroC}
\!\!\!\!.
\end{equation}
It is analogous to Eq.\,\eqref{CanonC}, but with the average thermal energy $\ave{E}_\beta$ replaced by the most probable energy $\max{E}_\beta$.
Note that the equations $\beta^{\rm mic}(E)\=\beta$ (equivalent to $E\=\max{E}_\beta$) and $E=\ave{E}_\beta$ are referred to as microcanonical and canonical caloric equations, respectively, while their solution in the plane $E\times\beta$ are called microcanonical and canonical caloric curves.

It is clear that the the microcanonical and canonical heat capacities are directly derived from the slopes of the respective caloric curves.
In standard thermodynamics, where one expects a convergence of $\ave{E}_\beta/N$ and $\max{E}_\beta/N$ for large $N$, both canonical and microcanonical heat capacities become identical in the thermodynamic limit.
However, in the $N\to\infty$ limit of low-$f$ systems the reality can be different.
In particular, Eq.\,\eqref{Microb} can yield a multiple solution in energy, or in contrary no solution within a certain domain of $\beta$.
This indicates more complex forms of the thermal energy distribution.
In such a situation, the microcanonical heat capacity is not well defined, e.g., has several branches corresponding to the multiple extremes of $w_{\beta}(E)$.

It is a simple exercise to calculate both canonical and microcanonical heat capacities for power-law Hamiltonians.
In this case, the microcanonical heat capacity has no ambiguity but differs from the canonical result.
Consider either separable or rotationally symmetric Hamiltonian forms,
\begin{subequations}
\label{Hpow}
\begin{align}
\label{Hsep}
H_{\rm sep}(\vecb{x})&=\sum_{i=1}^{f} \left(a_i|p_i|^{J_i}+b_i|q_i|^{I_i}\right)
\,,\\
H_{\rm rot}(\vecb{x})&=a|\vecb{p}|^{J}+b|\vecb{q}|^{I}
\label{Hrot}
\,,
\end{align}
\end{subequations}
where $a_i,b_i$ and $a,b$ are positive coefficients and $J_i,I_i$ and $J,I$ arbitrary (not necessarily integer) positive powers.
The semiclassical partition function and level density in both cases can be evaluated analytically, yielding power-law dependences $Z\propto \beta^{-M}$ and $\rho\propto E^{M-1}$, where the power constants $M$ for both Hamiltonian forms \eqref{Hsep} and \eqref{Hrot} are given by
\begin{subequations}
\label{Mpow}
\begin{align}
\label{Msep}
M_{\rm sep}&=\sum_{i=1}^{f}\left(\frac{1}{J_i}+\frac{1}{I_i}\right)
\,,\\
M_{\rm rot}&=\frac{f}{J}+\frac{f}{I}
\label{Mrot}
\,.
\end{align}
\end{subequations}
The canonical and microcanonical heat capacities for the respective cases read as 
\begin{subequations}
\label{Cpow}
\begin{align}
\label{Csep}
C^{\rm can}_{\rm sep}&=M_{\rm sep}\,,\quad
C^{\rm mic}_{\rm sep}=M_{\rm sep}-1
\,,\\
C^{\rm can}_{\rm rot}&=M_{\rm rot}\,,\quad
C^{\rm mic}_{\rm rot}=M_{\rm rot}-1
\label{Crot}
\,.
\end{align}
\end{subequations}

We stress that the microcanonical thermodynamics is in the present case defined only if $M\>1$, otherwise the level density is a non-increasing function of energy and Eq.\,\eqref{Microb} has no solution $\max{E}_\beta$ for $\beta\!\geq\!0$.
For purely quadratic kinetic terms, $J_i\=2$, this condition is not satisfied in $f\=1$ systems with quadratic or steeper potentials, $I_i\geq 2$ (for instance, quartic and hard-wall 1-dimensional potentials yield decreasing level densities $\rho\propto E^{-1/4}$ and $E^{-1/2}$, respectively).
On the other hand, the canonical results are conceivable for any $M\>0$. 

As seen from Eq.\,\eqref{Cpow}, the canonical and microcanonical heat capacities for power-law Hamiltonians satisfy a relation $C^{\rm mic}=C^{\rm can}\-1$.
The reason is that the thermal energy distribution \eqref{Ether} with $\rho\propto E^{M-1}$ has the average energy $\ave{E}_\beta$ systematically larger than the most probable energy $\max{E}_\beta$, and this asymmetry increases linearly with temperature.
For low \DoF\ numbers (small values of $M$), the difference is significant. 
However, as the \DoF\ number increases, both canonical and microcanonical specific heats calculated per a single \DoF\ become indistinguishable, reaching equality $C^{\rm can}/f=C^{\rm mic}/f$ in the limit $f\to\infty$.

\section{Heat capacity in presence of \textsc{esqpt}s}
\label{sec:EsqptCgen}

The \ESQPT\ represents an $N\!\to\!\infty$ singularity in the level density of a finite-$f$ quantum system caused by a stationary point of the corresponding classical Hamiltonian.
The singularity affects also many other observables, e.g., the \uvo{flow} of energy levels with a running control parameter, the form of eigenstates and evolution of some expectation values in the critical energy region, response of the system to external probes and so on
\cite{Cej06,Cap08,Cej08,Rel08,Fer11,Bra13,Lar13,Die13,Bas14,Str14,Str15,Cej15,Kop15,Pue15,San15,San16,Str16}.
Here we will investigate some consequences on thermal behavior.

A classification of the generic \ESQPTs\ for system with an arbitrary \DoF\ number $f$ was derived in Ref.\,\cite{Str16}.
It was shown that a non-degenerate stationary point (\NDSP) with rank $r$ of the Hamiltonian function $H(\vecb{x})$ in the $2f$-dimensional phase space (that is, a locally quadratic stationary point with $r$ negative eigenvalues of the Hessian matrix $\partial^2 H/\partial x_i\partial x_j$)
causes a singularity in the $(f\-1)$th derivative of the level density.
The singularity has a form
\begin{equation}
\frac{\partial^{f-1}\rho}{\partial E^{f-1}}\propto
\left\{\begin{array}{ll}
(-)^{r/2}\Theta(E\-E_{\rm c}) & r\={\rm even,}\\ 
(-)^{(r+1)/2}\ln|E\-E_{\rm c}|& r\={\rm odd,} 
\end{array}\right.
\label{dender}
\end{equation}
where $\Theta$ stands for a step function, $\Theta(x)\=1$ or 0 for $x\!\geq\!0$ or $x\<0$, respectively, and $E_{\rm c}$ denotes the critical energy at which the \ESQPT\ occurs.
We observe that at the energy of an \NDSP, the derivative $\partial^{f-1}\rho/\partial E^{f-1}$ shows either a jump (if the index $r$ is even) or a logarithmic divergence (if $r$ is odd).
For degenerate (flatter than quadratic) stationary points, the character of the non-analyticity of $\rho$ can be different (possibly affecting lower derivatives), but these cases are rare (we will see an example below).

The \ESQPT\ singularities have a direct impact on the microcanonical thermodynamics of the system, and indirectly modify also the canonical thermodynamics.
Determination of the microcanonical quantities relies on uniqueness of the solution of the microcanonical caloric equation, that is on the monotonicity of the microcanonical (inverse) temperature \eqref{Microb} as a function of energy.
This is commonly violated in particular for low \DoF\ numbers.
For $f\=1$, the presence of any \ESQPT\ implies a discontinuous or divergent behavior of the level density, so the microcanonical caloric curve cannot even be defined.
For $f\=2$, an \NDSP-related \ESQPT\ leads to a jump or logarithmic divergence of $\beta^{\rm mic}(E)$, so the microcanonical caloric equation has multiple or no solutions in wide temperature domains.
This was discussed in Ref.\,\cite{Str14}, so in the following we focus on the $f\geq 3$ cases.

A unique solution of the microcanonical caloric equation can fail even for $f\=3$.
Considering the \NDSP-related singularities, this is so near an \ESQPT\ with $r\=1,5$ (the microcanonical caloric curve in $E\times\beta \equiv x\times y$ has an upward vertical tangent at the critical point), and it may be so near an \ESQPT\ with $r\=0,4$ (here the tangent makes an upward jump).
Both these singularities can give rise to a pair of adjacent local extremes of the microcanonical caloric curve.
As a consequence, $C^{\rm mic}$ in certain intervals of $\beta$ has three branches (two positive and one negative) and may exhibit some divergences (located at the edges of the intervals). 
An example will be discussed in Sec.\,\ref{sec:EsqptCnum}.

For $f\geq 4$, the \NDSP-related \ESQPTs\ do not any more create any upward pointing segments of the microcanonical caloric curve, so the solution of the caloric equation is unique.
Nevertheless, either form of non-analyticity \eqref{dender} in the level density causes an associated singularity in the corresponding derivative of the microcanonical heat capacity.
Indeed, using Eq.\,\eqref{MicroC}, one can express the $n$th derivative of $C^{\rm mic}$ in $\beta$ through the derivatives of $\rho$ in $E$:
\begin{equation}
\frac{\partial^n C^{\rm mic}}{\partial\beta^n}\=
\frac{(-C^{\rm mic})^{n+2}}{\beta^{2n+2}}\!\left(\frac{1}{\rho}\frac{\partial^{n+2}\rho}{\partial E^{n+2}}\right)_{\!\!\!E=\max{E}_\beta}
\!\!\!\!\!\! +R
\label{CMder}
\end{equation}
($n\geq 1$).
Here we explicitly write only the term that contains the highest derivative of $\rho$, while $R$ collects all the remaining terms, which contain only derivatives $\partial^k\rho/\partial E^k$ of lower orders $k\<n\+2$.
Assuming a discontinuity (divergence) in the $(f\-1)$th derivative of $\rho$, as follows from Eq.\,\eqref{dender} for any \NDSP-related \ESQPT, we expect the same type of non-analyticity affecting the $(f\-3)$th derivative of the microcanonical heat capacity $C^{\rm mic}$.
Examples of this behavior will be shown in Sec.\,\ref{sec:EsqptCnum}.

Let us turn now to the canonical heat capacity $C^{\rm can}$.
As follows from Eqs.\,\eqref{CanonC} and \eqref{CanonC1}, it must be a smooth, regular function of temperature, irrespective of the presence of \ESQPTs\ and an actual size of the \DoF\ number.
However, for systems that have complicated energy landscapes in the phase space (with minima, maxima and saddles, hence \ESQPTs) the canonical heat capacity typically exhibits a non-trivial dependence on temperature.

The location of increasing and decreasing segments of $C^{\rm can}$ as a function of $\beta$ is determined by an interplay of dispersion and asymmetry (skewness) of the thermal energy distribution \eqref{Ether}.
Indeed, defining the third moment of a quantity $W$ in the thermal ensemble,
\begin{align}
\ske{W}_\beta\equiv\left\langle(W\!\-\ave{W}_\beta)^3\right\rangle_\beta=\qquad\qquad
\nonumber\\
\ave{W^3}_\beta-3\ave{W^2}_\beta\ave{W}_\beta+2\ave{W}_\beta^3
\label{skew}
\,,
\end{align}
one can write a simple expression for the first derivative of the canonical heat capacity
\begin{equation}
\frac{dC^{\rm can}}{d\beta}=2\beta\dis{E}_\beta-\beta^2\ske{E}_\beta
\label{Cder}
\,.
\end{equation}
The asymmetry \eqref{skew} can in general be both positive or negative (more weight of the distribution above or below the average $\ave{W}_\beta$, respectively), but only the case $\ske{E}_\beta\>0$ may give rise to a non-monotonous temperature dependence of $C^{\rm can}$.
Every time when $\beta^3\ske{E}_\beta/2$ crosses $\beta^2\dis{E}_\beta\=C^{\rm can}$, the canonical heat capacity has a maximum, minimum or inflection.
The number and positions of these points depend sensitively on the specific shape of the Hamiltonian function $H(\vecb{x})$.

\begin{figure}[t]
\begin{center}\includegraphics[width=0.85\linewidth]{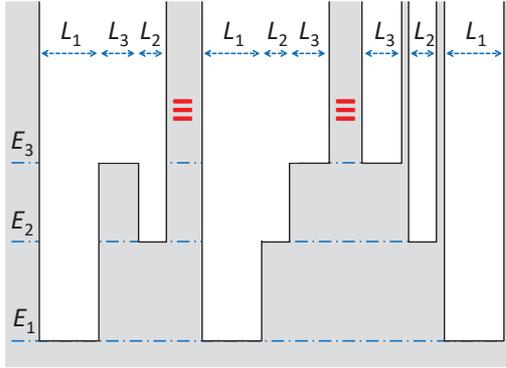}\end{center}
\caption{Examples of square-well potentials \eqref{wells} with equivalent semiclassical thermodynamic properties. Gray areas represent unreachable regions.}
\label{F_wells}
\end{figure}

To illustrate these matters on a toy example, let us consider a Hamiltonian of the form \eqref{Hstand} with a separable potential $V(\vecb{q})=\sum_{i=1}^{f}V_i(q_i)$.
The first component $V_1(q_1)$ is a multi-plateau infinite-square-well potential,
\begin{equation}
V_1=\left\{\begin{array}{ll}
E_1 & {\rm for\ } q\in[q_1^{\llcorner},q_1^{\lrcorner}),\\
\vdots & \\
E_m & {\rm for\ } q\in[q_m^{\llcorner},q_m^{\lrcorner}),\\
\infty & {\rm otherwise}
\end{array}\right.
\label{wells}
\end{equation}
where all intervals $[q_k^{\llcorner},q_k^{\lrcorner})$ with $k\=1,...,m$ are disjunct and have lengths $(q_k^{\lrcorner}\-q_k^{\llcorner})\equiv L_k$.
The remaining components $V_2(q_2),\dots V_f(q_f)$ are harmonic-oscillator potentials.
It should be stressed that thermodynamic results discussed below are independent of whether the intervals in Eq.\,\eqref{wells} are connected or not, so it does not make any difference whether the potential $V_1$ is a single infinite well with several bottom plateaus or a system of independent wells (or a combination of both).
The results are also independent of the ordering of individual energy plateaus $E_k$, so for example all configurations depicted in Fig.\,\ref{F_wells} are entirely equivalent.

\begin{figure}[t]
\begin{center}
\includegraphics[width=0.9\linewidth]{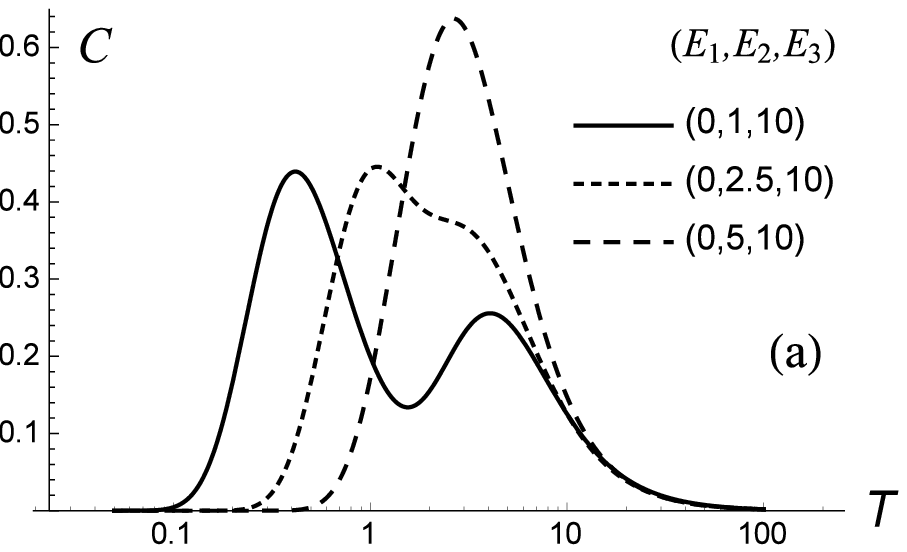}
\includegraphics[width=0.9\linewidth]{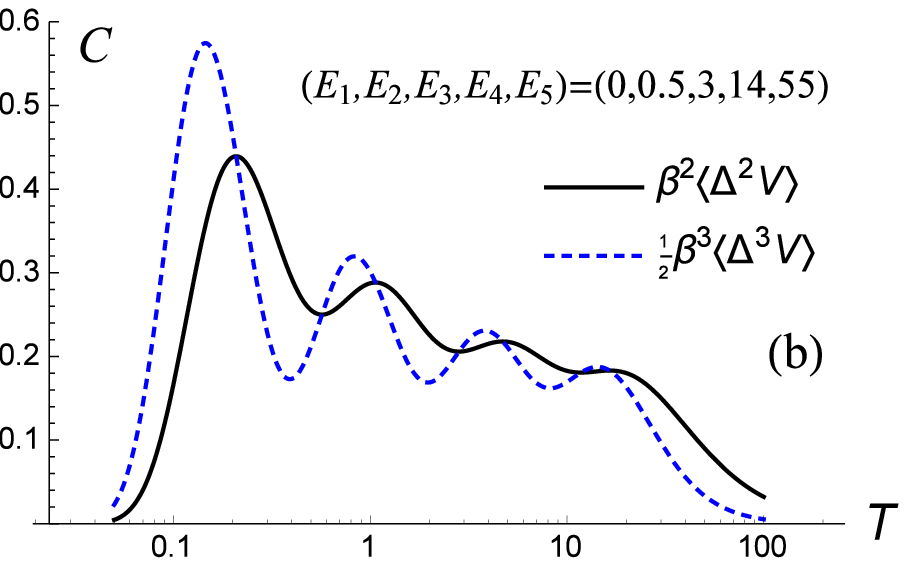}
\end{center}
\caption{The potential component of the canonical heat capacity for potentials of the form \eqref{wells}. Panel (a): The heat capacity for three choices of a 3-plateau potential (plateau energies $E_k$ given in the legend, $L_k\!=\!1$). Panel (b): The same for a single choice of a 5-plateau potential ($E_k$ indicated, $L_k\!=\!1$) along with a scaled asymmetry of the potential energy distribution.}
\label{F_wellheat}
\end{figure}

Almost everything related to the above-introduced model can be calculated by hand.
Note that instead of non-degenerate stationary points we now deal with an infinitely degenerate case, since inside the well(s) the derivatives with respect to $q_1$ vanish up to infinite order.
Nevertheless, the results will be analogous to those obtained for the \NDSP\ cases.
The level density of the $(f\-1)$-dimensional oscillator is given by $\rho_{\rm osc}\propto E_{\rm osc}^{f-2}$, while that of the 1-dimensional potential well system reads as $\rho_{\rm well}\propto\sum_{k=1}^{m}L_k\Theta(E_{\rm well}\-E_k)/\sqrt{E_{\rm well}\-E_k}$ (we are using the semiclassical approximation).
The total level density of the whole system is given by a convolution of the two partial level densities with $E\=E_{\rm well}\+E_{\rm osc}$ \cite{Str15}, yielding:
\begin{equation}
\rho(E)\propto\sum_{k=1}^{m}L_k\Theta(E\-E_k)(E-E_k)^{f-3/2}
\label{wellden}
\,.
\end{equation}
The $(f\-1)$th derivative of this function diverges at the energies $E_k$ corresponding to individual potential plateaus, which therefore become critical points of \ESQPTs\ of a non-\NDSP\ type.
This behavior is partly analogous to Eq.\,\eqref{dender}, except that the divergence is not logarithmic but $\propto\Theta(E\-E_{\rm c})/\sqrt{E\-E_{\rm c}}$, with $E_{\rm c}\!\equiv\!E_k$.

Following the directions given above, one would be able to determine the microcanonical heat capacity $C^{\rm mic}$ and to analyze the singularities in its $(f\-3)$th derivative connected with the critical points $E_k$.
It is also very easy to calculate the canonical quantities.
Thanks to separability of the system, the partition function factorizes and the canonical heat capacity $C^{\rm can}$ becomes a sum of terms corresponding to individual degrees of freedom.
The oscillators contribute by an overall constant $(f\-1)$ and the kinetic term of the first \DoF\ by an additional step-up by 1/2.
The only non-trivial contribution results from the second term of Eq.\,\eqref{CanonC2}, but it can also be calculated analytically.
Note that if all plateau energies are the same, $E_k\=E_0\ \forall k$, the contribution of the potential component $V_1$ is identically zero (the semiclassical thermal energy distribution for any temperature is fully localized at the bottom energy $E_0$, where the $f\=1$ level density diverges).
The situation becomes more interesting if the plateau energies are diversified.
Some examples of heat capacities for such potentials are shown in Fig.\,\ref{F_wellheat}. 

In the upper panel of Fig.\,\ref{F_wellheat} we present the potential component $\beta^2\dis{V_1}_\beta=C^{\rm can}\-f\+1/2$ of the canonical heat capacity as a function of temperature $T$ for the potential \eqref{wells} with 3 plateaus.
Individual curves correspond to various choices of the plateau energies $E_k$, the widths $L_k$ of all plateaus are the same.
All dependences start from a zero value at $T\=0$, where only the lowest plateau is populated, but as soon as the thermal energy distribution \eqref{Ether} spreads across the energies of higher plateaus, the heat capacity increases.
The dependence is non-monotonous and has 1 or 2 local maxima.
In the lower panel of Fig.\,\ref{F_wellheat} we present the potential component of $C^{\rm can}$ for a well with 5 plateaus, and in that case one can observe a sequence of up to 4 local maxima.
The peaks in this sequence correspond to the temperature domains in which the population rate of individual plateaus with successive energies is maximized.
If two or more plateau energies get closer to each other (i.e., of the same magnitude, their ratio still being as large as $\sim 3$), the peaks in the heat capacity merge, reducing complexity of the dependence.
For $T\to\infty$, the potential component of $C^{\rm can}$ returns back to the zero value.

The lower panel of Fig.\,\ref{F_wellheat} shows also the curve $\beta^3\ske{V}_\beta/2$ (the dashed line), whose intersections with $\beta^2\dis{V}_\beta$ indicate local extremes of the latter, cf.\,Eq.\,\eqref{Cder}.
We see that the temperature dependence of the asymmetry of the potential-energy distribution is similar to that of the potential-energy dispersion, but with the maxima and minima scaled and shifted to lower temperatures.
The form of this dependence is hard to anticipate, nevertheless, it is a key for the behavior of the canonical heat capacity.

\section{Numerical examples}
\label{sec:EsqptCnum}

\begin{figure}[t]
\includegraphics[width=\linewidth]{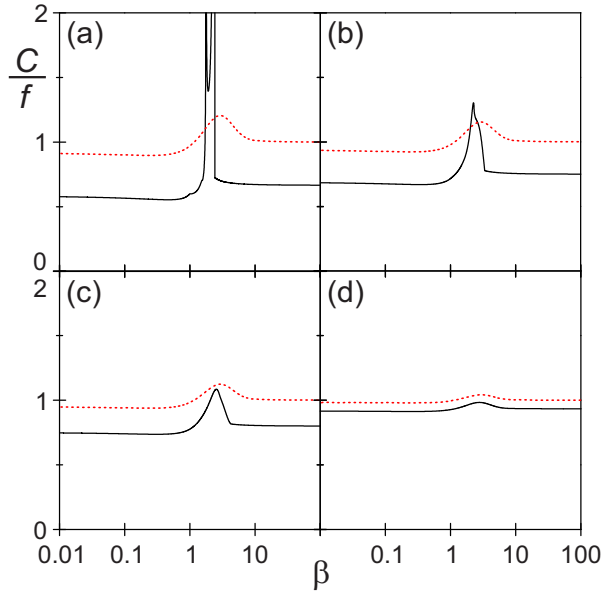}
\caption{The microcanonical and canonical heat capacity per \DoF\ (full and dashed curves, respectively) for a separable system composed of a single quartic oscillator with a double-well potential and $(f\!-\!1)$ harmonic oscillators. The panels from (a) to (d) correspond to $f=3,4,5,15$, respectively.}
\label{F_cusosc}
\end{figure}

\begin{figure}[t]
\begin{center}
\includegraphics[width=\linewidth]{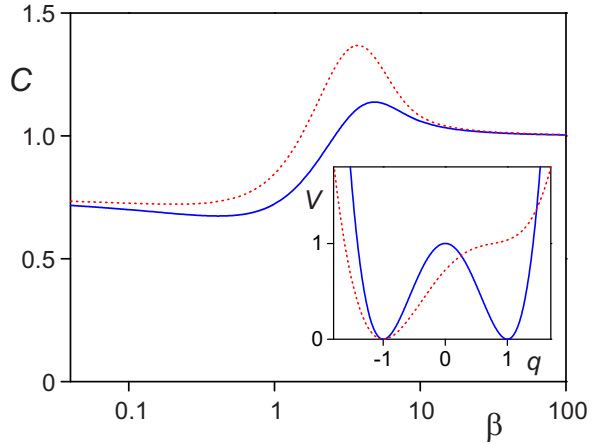}
\end{center}
\caption{(Color online) The canonical heat capacity for two $f\!=\!1$ systems with potentials drawn in the inset; the parameters $(a,b,c)$ from Eq.\,\eqref{quart} are $(0,-2,1)$ for the full curve and $(0.52,-0.52,0.26)$ for the dashed curve (the dashed potential is shifted to fit with the full one).}
\label{F_double}
\end{figure}

The above multi-plateau toy model provides useful intuition for understanding a more realistic model discussed in this section.
As in the previous case, we again consider a Hamiltonian of type \eqref{Hstand} with a separable potential $V=\sum_{i=1}^f V_i$, but individual terms now represent mixed linear-quadratic-quartic 1-dimensional potentials of the form
\begin{equation}
V_i=v_{i}+a_iq_i+b_iq_i^2+c_iq_i^4
\label{quart}
\,,
\end{equation}
where $v_i$, $a_i,b_i$ and $c_i\!\geq\!0$ are adjustable parameters.
If $c_i\>0$, the potential for $b_i\<0$ and $|a_i|\<\sqrt{8|b_i|^3/27c_i}\!\equiv\! a_{\rm W}$ represents a double well, while that for $b_i\!\geq\!0$ or $|a_i|\!\geq\!a_{\rm W}$ is a single well.
In both these cases, the potential has a quartic asymptotics and it is asymmetric unless $a_i\=0$.
On the other hand, if $c_i\=0$ and $b_i\>0$, the potential is quadratic, for $a_i\=0$ coinciding with an ordinary harmonic oscillator.
The constant $v_i$ is always chosen so that the potential has its absolute minimum at the value $V_i\=0$.
If not specified otherwise, we choose $c_i\=1,b_i\=-2$ for quartic potentials and $b_i\=1,a_i\=0$ for quadratic potentials.
Note that the coefficients are chosen independently for each \DoF, so we allow for situations in which only one of the potentials $V_i$ has the double-well form while the others are harmonic oscillators, as well as those in which each $V_i$ contains a non-trivial quartic dependence.

\begin{figure*}[t]
\begin{center}
\includegraphics[width=0.63\linewidth]{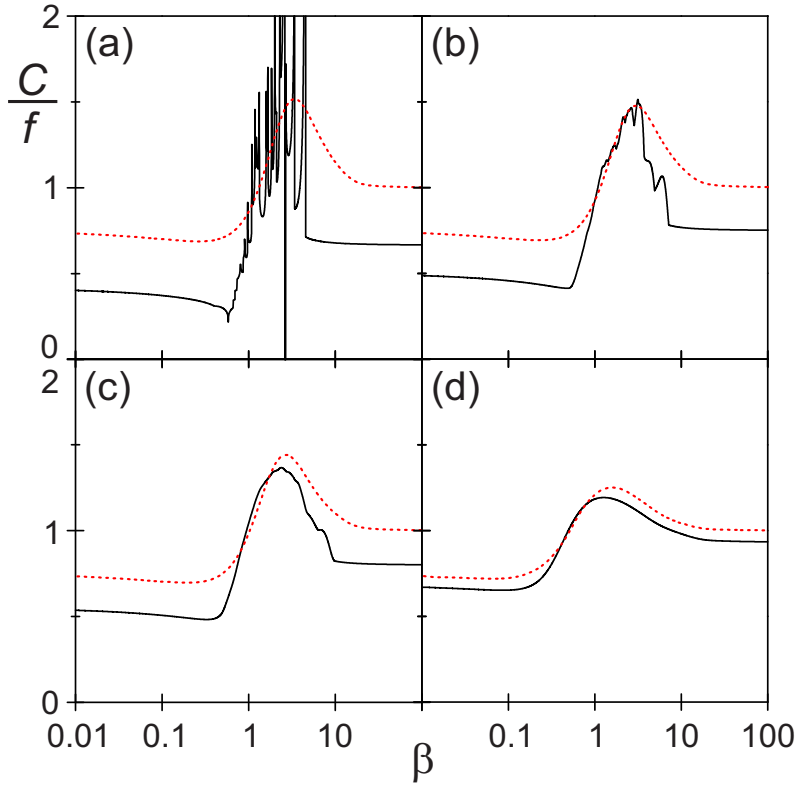}
\end{center}
\caption{The same as in Fig.\,\ref{F_cusosc} but for a separable system of $f$ quartic oscillators.}
\label{F_cuscus}
\end{figure*}

The first type of situation is investigated in Fig.\,\ref{F_cusosc}. 
It corresponds to systems with $f\=3,4,5$ and 15 [panels from (a) to (d)], all having just 3 stationary points generated by the double-well potential $V_i$ in only one \DoF. 
More specifically, the overall potential $V$ is composed of a single quartic oscillator with $a_i\=0.5$ for $i\=1$, and $(f\-1)$ replicas of the same harmonic oscillator for $i\=2,...,f$.
Fig.\,\ref{F_cusosc} compares the canonical and microcanonical heat capacities per \DoF, both as a function of $\beta$ (so the horizontal axis is inverted in comparison with the dependence on $T$, cf.\,Fig.\,\ref{F_wellheat}).
We observe that the canonical curve $C^{\rm can}/f$ is smooth and has a single maximum.
The limiting values, namely $(1\-1/4f)$ for $\beta$ small and $1$ for $\beta$ large, follow from a prevailingly quadratic behavior of the potential at low energies near the minimum (at low temperatures) and a mixed [$(f\-1)\times$\,quadratic, $1\times$\,quartic] dependence at large energies (large temperatures).
Also the microcanonical curve $C^{\rm mic}/f$ has a peak-like behavior, but it moreover exhibits 3 points of non-analyticity. 
They result from the level-density singularities caused by the stationary points and, as expected, are most apparent for low \DoF\ numbers. 
For $f\=3$, we observe divergences of $C^{\rm mic}$ connected with singular segments of the microcanonical caloric curve (see Sec.\,\ref{sec:EsqptCgen}). 
As $f$ increases, the non-analyticities become virtually invisible (moving to higher derivatives), and the canonical and microcanonical curves get close to each other, both approaching the limit $C^{\rm can}/f\=C^{\rm mic}/f\=1$.

Non-monotonous dependences of $C^{\rm can}$ in Fig.\,\ref{F_cusosc} are similar to those from Fig.\,\ref{F_wellheat}.
The reason for this behavior is a gradual population of states located in individual potential wells and of the higher-energy states centered in the barrier region.
As in the multi-plateau case, the results are partly independent of a spatial redistribution of the potential strength.
Fig.\,\ref{F_double} compares an $f\=1$ canonical heat capacity (both kinetic and potential terms included) for two potential forms, namely a degenerate double well and a single well with a \uvo{plateau} (both potentials are drawn in the inset).\footnote
{
The canonical partition function for a degenerate double-well oscillator $V=x^4-2x^2$ is given by an analytic expression 
$Z(\beta)=\frac{\pi}{2}\,e^{\beta/2}\left[I_{-1/4}\left(\tfrac{\beta}{2}\right)+I_{+1/4}\left(\tfrac{\beta}{2}\right)\right]$,
where $I_k(x)$ is a modified Bessel function of the first kind, while that for the single well potential is calculated numerically. 
}
One may imagine that the wide minimum and the plateau of the second potential result, respectively, from merging the two minima of the first potential and moving the barrier to an edge.
Indeed, as seen in the main panel, both potentials yield qualitatively similar dependences of $C^{\rm can}$.

\begin{figure}[ht!]
\includegraphics[width=\linewidth]{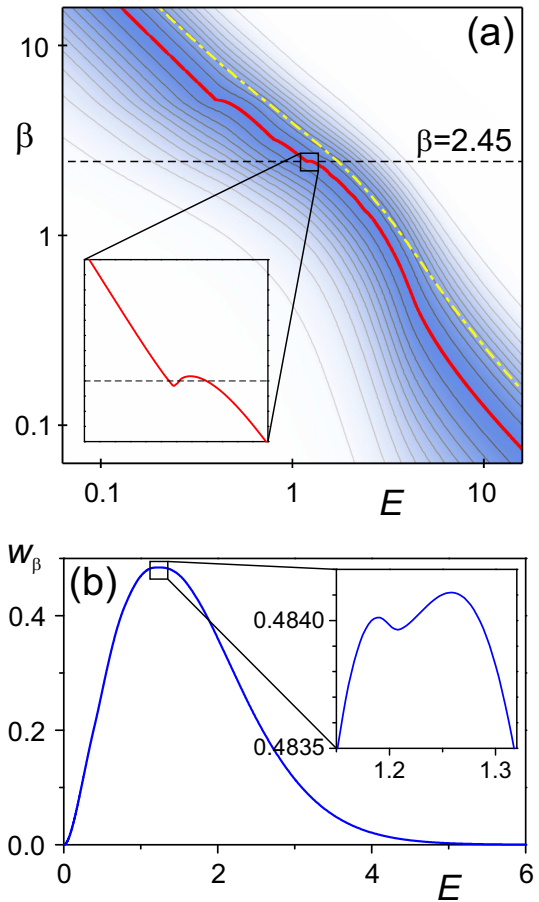}
\caption{(Color online) Thermal energy distributions for a separable $f\!=\!3$ system composed of 3 double-well potentials. Panel (a): Distribution $w_\beta(E)$ as a function of $E\times\beta$; the darker band is the region with large values (the maximum normalized to unity). The microcanonical and canonical caloric curves, respectively, are drawn by the thicker full line (red online) and the dot-dashed line (yellow online). Panel (b): Thermal energy distribution at $\beta\!=\!2.45$ with the region of bimodality (expanded in the inset) corresponding to the kink of the caloric curve (expanded in the inset of the upper panel).}
\label{F_sward}
\end{figure}

The second type of a separable system based on the above potential---that with a large number of stationary points resulting from a combination of individual quartic oscillators in all \DoFs---is studied in Fig.\,\ref{F_cuscus}.
The figure shows the dependences of $C^{\rm can}/f$ and $C^{\rm mic}/f$ on $\beta$ for systems (again with $f\=3,4,5,15$) composed of $i\=1,...,f$ potentials $V_i$ with $a_i=i/5$.
Note that a direct sum of $n$ potentials of a double-well form (with 3 stationary points) and $(f\-n)$ potentials with a single-well form (1 stationary point) leads to a potential $V$ with an overall number of $3^n$ stationary points.
In the present parametrization, the potentials $V_i$ have a double-well form for $i\<8$, so the numbers of stationary points for $f\=3,4,5$ and 15 read as $27,81,243$ and 2187, respectively.
As in the previous case, the effect of stationary points in the microcanonical curves is strongest for low \DoF\ numbers.
For $f\=3$, e.g., we again observe divergences, but even a short negative branch of $C^{\rm mic}$; the discussion of this case follows in connection with Fig.\,\ref{F_sward}. 
For growing $f$, the canonical and microcanonical curves increasingly overlap.
Note that the present high-temperature limit $C^{\rm can}/f\=3/4$ results from purely quartic asymptotics of the potential.

In Fig.\,\ref{F_sward} we show thermal energy distributions $w_\beta(E)$ from Eq.\,\eqref{Ether} for an $f\=3$ system combining three double-well potentials.
The parameters are the same as in Fig.\,\ref{F_cuscus}(a). 
The thermal energy distribution for a single value of $\beta$ is depicted in the lower panel, while a collection of such distributions for $\beta$ in a wide interval is visualized in the upper panel by a combined contour--density plot in the $E\times\beta$ plane (the lower panel is just a cut of the upper panel along the dashed horizontal line).
Note that the energy distribution for each $\beta$ is normalized to a fixed value at the maximum.
We also show in the upper panel the canonical and microcanonical caloric curves (full and dot-dashed lines, respectively). 
These are obtained as solutions of the respective caloric equation.
While the canonical curve $E\=\ave{E}_\beta$ is smooth and monotonous (decreasing with $\beta$), the microcanonical curve $E\=\max{E}_\beta$ has some kinks. 
One of them is expanded in the inset.
Since the microcanonical caloric curve represents a locus of points where the condition $\partial w_\beta/\partial E\=0$ is satisfied, the kinks generate some shallow bimodalities of the $w_\beta(E)$ distribution---see the inset in the lower panel.
Precisely these structures are responsible for multi-valued dependences of $C^{\rm mic}$ in the corresponding narrow intervals of $\beta$.
While the two decreasing segments of the microcanonical caloric curve generate positive branches of $C^{\rm mic}$, the increasing segment results in a negative branch.
In addition, the local extremes of the microcanonical caloric curve can give rise to divergences of $C^{\rm mic}$ at the edges of the anomalous intervals (this nevertheless requires differentiability of the caloric curve at the respective extremal points).
Numerical specimens of such pathologies were seen in Fig.\,\ref{F_cuscus}(a).

\section{Conclusions}
\label{sec:Conc}

We have studied non-standard thermodynamics of quantum systems with an effective \DoF\ number $f$ independent from the size parameter $N$.
The study is relevant for various models of collective dynamics, in which the non-collective \DoFs\ become inactive, frozen in a given physical context or experimental setup.
Examples can be found within the family of interacting boson models used in molecular ($f\=2,3$) and nuclear ($f\!\geq\!5$) physics (possible coupling of models in both molecular and nuclear context leads to a multiplication of the \DoF\ numbers), see Ref.\,\cite{Fra05}.
Applications in the physics of synthetic quantum systems can be based on the Lipkin- and Dicke-like models \cite{Kee14} describing for instance $N_{\rm s}$-site arrays of trapped 2-level atoms interacting with each other and possibly with an $N_{\rm m}$-mode bosonic field ($f\=N_{\rm s}\+N_{\rm m}$).

We have shown that the low-$f$ collective systems exhibit considerable differences between the canonical and microcanonical heat capacities $C^{\rm can}$ and $C^{\rm mic}$ even in the $N\to\infty$ limit. 
These differences are particularly dramatic for systems with \ESQPTs, in which $C^{\rm mic}$ shows non-analyticities at the \ESQPT\ critical energies (it cannot be properly defined for $f\!\leq\!3$), while $C^{\rm can}$ is a smooth (though non-monotonous) function of temperature.
The microcanonical singularities are unavoidable since they are rooted in the \ESQPT\ non-analyticities of the level density.
On the other hand, the canonical quantities must be smooth as long as the number of \DoFs\ remains finite. 
Both canonical and microcanonical approaches converge for $f\to\infty$ when $C^{\rm can}$ and $C^{\rm mic}$ grow roughly linearly with $f$ while $(C^{\rm can}\-C^{\rm mic})/f\to 0$.

The above findings demonstrate the fact that for collective systems, the infinite-size limit $N\!\to\!\infty$ does not coincide with the true thermodynamic limit.
The latter is achieved only if $f\!\to\!\infty$.
Smoothness of the canonical heat capacity in finite-$f$ systems indicates that the \ESQPTs---which (at least in their generic forms) can be observed only in finite \DoF\ numbers---are not associated with any kinds of thermal phase transition based on canonical thermodynamics.
This is in accord with the previous studies of the Dicke model \cite{Bas16,Rel16}.
The \ESQPTs\ and thermal phase transitions can therefore be considered as complementary forms of criticality in collective systems in the sense that they exist in the mutually \uvo{incompatible}\ asymptotic regimes of $N$ and $f$, respectively.

The thermodynamic signatures of \ESQPTs, as described in this work, may serve as a useful hint for future identification of \ESQPTs\ in specific systems.
Directly associated with the singularities of the microcanonical heat capacity are the irregular segments of the microcanonical caloric curve. 
They may affect thermal properties of isolated systems (in which thermalization is due to other means than a classical heat bath), although the survival of these relatively tiny effects on the background of fluctuations is not guaranteed.
However, \ESQPTs\ show up also in some non-trivial temperature dependences of the canonical heat capacity.
Its maxima/minima and increasing/decreasing segments result from competing populations of the states located around stationary points (or their remnants) of the Hamiltonian function in the phase space.
Such dependences can be seen as indicators of \ESQPTs\ that may appear nearby in the space of control parameters.

\section*{Acknowledgments}
%\vspace{-2mm}
The authors acknowledge discussions with T.\,Brandes.
This work was performed under the project no.\,P203-13-07117S of the Czech Science Foundation.

\end{document}